\documentclass[12pt]{article}

\usepackage{graphics,graphicx}
\usepackage[dvips]{epsfig}
\usepackage{eepic}
\usepackage{amsmath,amssymb,amsfonts}
\usepackage{color}
\long\def\commabs #1\commabsend{#1}
\long\def\commful #1\commfulend{}

\commful
\usepackage{times}
\commfulend

\commabs
\commabsend

\commful
\commfulend

\def\inline#1:{\par\vskip 7pt\noindent{\bf #1:}\hskip 10pt}

\commful

\commfulend
\commabs

\commabsend

\newtheorem{theorem}{Theorem}[section]

\newtheorem{definition}[theorem]{Definition}

\newtheorem{claim}[theorem]{Claim}
\newtheorem{lemma}[theorem]{Lemma}

\newtheorem{corollary}[theorem]{Corollary}

\newcommand{\qed}{\hfill $\Box$ \medbreak}
\newenvironment{proof}{\noindent {\bf Proof.}}{\qed}

\newcommand{\local}{{\cal LOCAL}}
\newcommand{\inp}{\mbox{\rm\bf x}}
\newcommand{\certif}{\mbox{\rm\bf y}}
\newcommand{\out}{\mbox{\rm out}}
\newcommand{\id}{\mbox{\rm Id}}
\newcommand{\dist}{\mbox{\rm dist}}

\newcommand{\nbnode}{\mbox{\tt \#n}}

\newcommand{\LD}{\mbox{\rm LD}}
\newcommand{\NLD}{\mbox{\rm NLD}}

\newcommand{\BPLD}{\mbox{\rm BPLD}}
\newcommand{\BPNLD}{\mbox{\rm BPNLD}}

\newcommand{\tree}{\mbox{\tt tree}}
\newcommand{\inpsize}{\mbox{\tt InpEqSize}}
\newcommand{\consensus}{\mbox{\tt Consensus}}
\newcommand{\mapcover}{\mbox{\tt MC}}
\newcommand{\containment}{\mbox{\tt containment}}
\newcommand{\cover}{\mbox{\tt cover}}
\newcommand{\leader}{\mbox{\tt Unique-Leader}}

\newcommand{\coloring}{\mbox{\tt Coloring}}

\newcommand{\mst}{\mbox{\tt MST}}
\newcommand{\mis}{\mbox{\tt MIS}}
\newcommand{\spanningtree}{\mbox{\tt SpanningTree}}

\def\cA{{\cal A}}
\def\cE{{\cal E}}
\def\cF{{\cal F}}
\def\cG{{\cal G}}
\def\cL{{\cal L}}

\def\cS{{\cal S}}

\newenvironment{smallitemize} {
  \begin{list}{$\bullet$} {\setlength{\parsep}{0pt}
\setlength{\itemsep}{0pt}} } { \end{list} }

\long\def\jump#1\finjump{}

\title{Local Distributed Decision
\thanks{Supported by a France-Israel cooperation grant 
(``Mutli-Computing'' project) 
from the France Ministry of Science and Israel Ministry of Science.}
}
\author{
Pierre Fraigniaud
\thanks{CNRS and University Paris Diderot, France. 
E-mail: {\tt \{pierre.fraigniaud,amos.korman\}@liafa.jussieu.fr}.
Supported by the ANR projects ALADDIN and PROSE and by the INRIA project GANG.}
\and Amos Korman$^\dag$
\and David Peleg
\thanks{
The Weizmann Institute of Science, Rehovot, Israel. 
E-mail: {\tt david.peleg@weizmann.ac.il}.}
}

\date{}

\begin{document}
\begin{titlepage}
\def\thepage{}
\maketitle

\begin{abstract}
A central theme in distributed network algorithms concerns understanding 
and coping with the issue of {\em locality}. 
Despite considerable progress, research efforts in this direction
have not yet resulted in a solid basis in the form of 
a fundamental computational complexity theory.  
Inspired by sequential complexity theory, we focus on a complexity theory
for \emph{distributed decision problems}. 
In the context of locality, solving a decision problem requires the processors
to independently inspect their local neighborhoods and then collectively decide
whether a given global input instance belongs to some specified language. 

We consider the standard $\cal{LOCAL}$ model of computation 
and define $\LD(t)$ (for {\em local decision}) as the class of decision problems 
that can be solved in $t$ number of communication rounds. 
We first study the intriguing question of whether 
randomization helps in local distributed computing, and to what extent.
Specifically, we define the corresponding randomized class $\BPLD(t,p,q)$,  containing languages for which there exists a randomized algorithm that runs in $t$ rounds and accepts correct instances with probability at least $p$ and rejects incorrect ones with probability at least $q$. 
We show that there exists a language that does not belong to $\LD(t)$ for any $t=o(n)$ but which belong for 
$\BPLD(0,p,q)$ for any $p,q\in (0,1]$ such that $p^2+q\leq 1$. On the other hand, we show that, restricted to hereditary languages,  
$\BPLD(t,p,q)=\LD(O(t))$, for any function $t$ and any $p,q\in (0,1]$ such that $p^2+q> 1$.

In addition, we investigate the impact of non-determinism on local decision, and establish some structural results inspired by classical computational complexity theory.   
Specifically, we show that non-determinism does help, but that this help is limited, as there exist languages that cannot be decided  non-deterministically. 
Perhaps surprisingly,  it turns out that it is the combination of randomization with non-determinism that enables to decide \emph{all} languages in constant time. 
Finally, we introduce the notion of  local reduction, and establish some completeness results.



\end{abstract}


\end{titlepage}

\section{Introduction}

\subsection{Motivation}

Distributed computing concerns a collection of processors which collaborate 
in order to achieve some global task. 
With time, two main disciplines have evolved in the field. 
One discipline deals with \emph{timing} issues, namely, 
uncertainties due to asynchrony (the fact that processors run 
at their own speed, and possibly crash), and the other concerns
\emph{topology} issues, namely, uncertainties due to locality constraints 
(the lack of knowledge about far away processors). Studies carried out 
by the distributed computing community within these two disciplines 
were to a large extent problem-driven. Indeed, several major problems 
considered in the literature concern coping with one of the two uncertainties. 
For instance, in the \emph{asynchrony-discipline}, 
Fischer, Lynch and Paterson~\cite{FLP85} proved that consensus cannot 
be achieved in the asynchronous model, even in the presence of a single fault, 
and in the \emph{locality-discipline}, Linial~\cite{L92} proved that 
$(\Delta+1)$-coloring cannot be achieved locally (i.e., in a constant number 
of communication rounds), even in the ring network.  

One of the significant achievements of the asynchrony-discipline 
was its success in establishing unifying theories in the flavor of 
computational complexity theory. Some central examples of such theories 
are failure detectors~\cite{CT96,CHT96} and the wait-free hierarchy 
(including Herlihy's hierarchy)~\cite{H91}. 
In contrast, despite considerable progress, the locality-discipline 
still suffers from the absence of a solid basis in the form of 
a fundamental computational complexity theory. Obviously, defining 
some common cost measures (e.g., time, message, memory, etc.) 
enables us to compare problems in terms of their relative cost. 
Still, from a computational complexity point of view, it is not clear 
how to relate the difficulty of problems in the locality-discipline. 
Specifically, if two problems have different kinds of outputs, 
it is not clear how to reduce one to the other, even if they cost the same. 

Inspired by sequential complexity theory, we focus on \emph{decision problems},
in which one is aiming at deciding whether a given global input instance 
belongs to some specified language. In the context of distributed computing, 
each processor must produce a boolean output, and the decision is defined by 
the conjunction of the processors' outputs, i.e., if the instance belongs 
to the language, then all processors must output~``yes'', and otherwise, 
at least one processor must output ``no''. Observe that decision problems 
provide a natural framework for tackling fault-tolerance: the processors 
have to collectively check whether the network is fault-free, and a node 
detecting a fault raises an alarm. In fact, many natural problems can be 
phrased as decision problems, like 
``is there a unique leader in the network?'' or ``is the network planar?''. 
Moreover, decision problems occur naturally when one is aiming at checking 
the validity of the output of a computational task, such as 
``is the produced coloring legal?'', or ``is the constructed subgraph an MST?''. 
Construction tasks such as exact or approximated solutions to problems like 
coloring, MST, spanner, MIS, maximum matching, etc., received enormous 
attention in the literature (see, e.g., 
\cite{BM09,KP_98,KW06, L92, L86,LPR09,LPP08, P08}), 
yet the corresponding decision problems have hardly been considered. 

The purpose of this paper is to investigate the nature of local decision 
problems. 
Decision problems seem to provide a promising approach to building up 
a distributed computational theory for the locality-discipline. 
Indeed, as we will show, one can define local reductions 
in the framework of decision problems, thus enabling the introduction of 
complexity classes and notions of completeness.

We consider the $\local$ model~\cite{Peleg:book}, which is a standard 
distributed computing model capturing the essence of locality. 
In this model, processors are woken up simultaneously, 
and computation proceeds in fault-free synchronous rounds during which 
every processor exchanges messages of unlimited size with its neighbors, 
and performs arbitrary computations on its data. 
Informally, let us define $\LD(t)$ (for local decision) as the class of 
decision problems that can be solved in $t$ number of communication 
rounds in the $\local$ 
\commful
model\footnote{
We find special interest in the case where $t$ represents a constant, but in general we view $t$ as a function of  the 
input graph.
We note that in the $\local$ model, every decidable decision problem 
can be solved in $n$ communication rounds, where $n$ denotes the number of nodes in the input graph.}. 
\commfulend
\commabs
model. (
We find special interest in the case where $t$ represents a constant, but in general we view $t$ as a function of  the 
input graph.
We note that in the $\local$ model, every decidable decision problem 
can be solved in $n$ communication rounds, where $n$ denotes the number of nodes in the input graph.)
\par
\commabsend
Some decision problems are trivially in $\LD(O(1))$ 
(e.g., ``is the given coloring a $(\Delta+1)$-coloring?'', 
``do the selected nodes form an MIS?'', etc.), while some others 
can easily be shown to be outside $\LD(t)$, for any $t=o(n)$ 
(e.g., ``is  the network planar?'', ``is there a unique leader?'', etc.). 
In contrast to the above examples, there are some languages for which 
it is not clear whether they belong to $\LD(t)$, even for $t=O(1)$. To elaborate on this, consider  
the particular case where it is required to decide whether the network 
belongs to some specified family $\cF$ of graphs. 
If this question can be decided in a constant number of communication rounds,
then this means, informally, that the family $\cF$ can somehow 
be characterized by relatively simple conditions. 
For example, a family $\cF$ of graphs that can be characterized as 
consisting of all graphs having no subgraph from $\cal C$, for some 
specified finite set $\cal C$ of finite subgraphs, is obviously in $\LD(O(1))$. 
However, the question of whether a family of graphs can be characterized 
as above is often non-trivial.
For example, characterizing cographs as precisely the graphs with no induced 
$P_4$, attributed to Seinsche~\cite{Seinsche74}, is not easy, and requires
nontrivial usage of modular decomposition.


The first question we address is whether and to what extent randomization helps.
For $p,q\in(0,1]$, 
define $\BPLD(t,p,q)$ as the class of all distributed languages that 
can be decided by a  randomized distributed algorithm that runs in 
$t$ number of communication rounds and produces correct answers 
on legal (respectively, illegal) instances with probability at least $p$ 
(resp., $q$). An interesting observation is that for $p$ and $q$ such that $p^2+q\leq 1$, we have $\LD(t)\subsetneq\BPLD(t,p,q)$. 
In fact,  for such $p$ and $q$, there exists a language $\cL\in \BPLD(0,p,q)$, such that $\cL\notin \LD(t)$, for any $t=o(n)$. 
To see why, consider the following $\leader$ language. 
The input is a graph where each node has a bit indicating whether it is a leader or not. An input is in the language $\leader$ if and only if
there is at most one leader in the graph.
Obviously, this language is not in $\LD(t)$, for any $t<n$. We claim it is in $\BPLD(0,p,q)$, for $p$ and $q$ such that $p^2+q\leq 1$. 
Indeed, for such $p$ and $q$, we can design the following simple randomized algorithm that runs in 0 time: every node which is 
not a leader says ``yes'' with probability 1, and every node which is a leader says ``yes'' with probability $p$. Clearly, if the graph has at most one leader
then all nodes say ``yes'' with probability at least $p$. On the other hand, if there are at least $k\geq 2$ leaders, at least one node says ``no'', with probability at least 
$1-p^k\geq 1-p^2\geq q$. 

It turns out that the aforementioned choice of $p$ and $q$ is not coincidental, and that $p^2+q=1$ is really the correct threshold. Indeed, we show that  
 $\leader\notin \BPLD(t,p,q)$, for any $t<n$, and any $p$ and $q$  such that $p^2+q>1$.
In fact, we show a much more general result, that is, we prove that if  $p^2+q>1$, then restricted to hereditary languages, $\BPLD(t,p,q)$ actually collapses into 
$\LD(O(t))$, for any $t$. 


In the second part of the paper, we investigate the impact of non-determinism on local decision, and establish some structural results inspired by classical computational complexity theory.   
Specifically, we show that non-determinism does help, but that this help is limited, as there exist languages that cannot be decided  non-deterministically. 
Perhaps surprisingly,  it turns out that it is the combination of randomization with non-determinism that enables to decide \emph{all} languages in constant time. 
Finally, we introduce the notion of  local reduction, and establish some completeness results.

\subsection{Our contributions}
\subsubsection{Impact of randomization}

We study the impact of randomization on local decision.
We  prove that if  $p^2+q>1$, then restricted to hereditary languages, $\BPLD(t,p,q)=\LD(O(t))$, for any function $t$. This, together with the observation that $\LD(t)\subsetneq \BPLD(t,p,q)$, for any $t=o(n)$,  may indicate that  $p^2+q=1$ serves as a sharp threshold for distinguishing  the deterministic case from the randomized one.

\subsubsection{Impact of non-determinism}

We first show that non-determinism helps local decision, i.e., we show that the class $\NLD(t)$ (cf. Section~\ref{subsec:NLD}) strictly contains $\LD(t)$. More precisely, we show that  there exists a language in $\NLD(O(1))$  which is not in $\LD(t)$ for every $t=o(n)$, where $n$ is the size of the input graph. Nevertheless, $\NLD(t)$ does not capture all (decidable) languages, for $t=o(n)$. Indeed  we show that  there exists a language not in $\NLD(t)$  for every $t=o(n)$. Specifically, this language is $\nbnode = \{(G,n) \mid |V(G)|=n\}.$ 

Perhaps surprisingly,  it turns out that it is the combination of randomization with non-determinism that enables to decide \emph{all} languages in constant time. Let $\BPNLD(O(1))= \BPNLD(O(1),p,q)$, for some constants $p$ and $q$ such that $p^2+q\leq1$. We prove that $\BPNLD(O(1))$ contains all languages. 
To sum up, $\LD(o(n))\subsetneq \NLD(O(1))\subset  \NLD(o(n)) \subsetneq \BPNLD(O(1))=\mbox{All}$.

Finally, we introduce the notion of many-one local reduction, and establish some completeness results. We show that there exits a problem, called \cover, which is, in a sense, the most difficult decision problem. That is we show that \cover\/ is $\BPNLD(O(1))$-complete.  (Interestingly, a small relaxation of \cover, called \containment, turns out to be $\NLD(O(1))$-complete).

\subsection{Related work}
\commful\parskip-0.05in\commfulend
Locality issues have been thoroughly studied in the literature, 
via the analysis of various construction problems, including 
$(\Delta+1)$-coloring and Maximal Independent Set (MIS)
~\cite{ABI,BM09,K09,KW06, L92, L86, PS96}, Minimum Spanning Tree (MST)
~\cite{E06,KP_98,PR_00}, Maximal Matching~\cite{HKP01}, 
Maximum Weighted Matching~\cite{LPR09,LPP08,WW}, Minimum Dominating Set
~\cite{KMW04,LOW08}, Spanners~\cite{DGPV09,EZ06,P08}, etc. 
For some problems (e.g., coloring~\cite{BM09,K09, PS96}), there are still 
large gaps between the best known results on specific families of graphs 
(e.g., bounded degree graphs) and on arbitrary graphs.
\commful\parskip0.0in\commfulend


The question of what can be computed in a constant number of communication 
rounds was investigated in the seminal work of Naor and Stockmeyer~\cite{NS93}.
In particular, that paper considers a subclass of $\LD(O(1))$, called LCL, 
which is essentially $\LD(O(1))$ restricted to languages involving graphs 
of constant maximum degree, and involving processor inputs taken from a set 
of constant size, and studies the question of how to compute 
in $O(1)$ rounds the constructive versions of decision problems in LCL. 
The paper provides some beautiful general results. In particular, the authors show 
that if there exists a randomized algorithm that 
constructs a solution for a problem in LCL in $O(1)$ rounds, 
then there is also a deterministic algorithm constructing a solution for 
this problem in $O(1)$ rounds. Unfortunately, the proof of this result 
relies heavily on the definition of LCL. Indeed, the constant bound constraints 
on the degrees and input sizes allow the authors to cleverly use 
Ramsey theory. It is thus not clear whether it is possible  to extend this result 
 to all languages in $\LD(O(1))$.

The question of whether randomization helps in decreasing the locality 
parameter of construction problems has been the focus of numerous  studies. 
To date, there exists evidence that, for some problems at least, 
randomization does not help. 
For instance, \cite{N91} proves this for 3-coloring the ring. 
In fact, for low degree graphs, the gaps between the efficiencies 
of the best known randomized and deterministic algorithms for problems like 
MIS, $(\Delta+1)$-coloring, and Maximal Matching are very small. 
On the other hand, 
for graphs of arbitrarily large degrees, there seem to be indications 
that randomization does help, at least in some cases. 
For instance, $(\Delta+1)$-coloring can be randomly computed in expected 
$O(\log n)$ communication rounds on $n$-node graphs~\cite{ABI,L86}, 
whereas the best known deterministic algorithm for this problem 
performs in $2^{O(\sqrt{\log n})}$ rounds~\cite{PS96}.  $(\Delta+1)$-coloring results whose performances  are measured also with respect to the maximum 
degree $\Delta$ illustrate this phenomena as well. Specifically,  \cite{SW10} shows that
 $(\Delta+1)$-coloring can be randomly computed in expected 
$O(\log \Delta +\sqrt{\log n})$ communication rounds
whereas the best known deterministic algorithm 
performs in $O(\Delta+\log^* n)$ rounds~\cite{BM09,K09}.


Recently, several results were established conserving decision problems in distributed computing.
For example,  \cite{DHKKNPPW} and \cite{KKP11} study specific
decision problems in the $\cal{CONGEST}$ 
\commful
model\footnote{In contrast to the $\cal{LOCAL}$ model, this model assumes 
that the message size is bounded by $O(\log n)$ bits, hence dealing with congestion 
is the main issue.}.
\commfulend
\commabs
model. (In contrast to the $\cal{LOCAL}$ model, this model assumes 
that the message size is bounded by $O(\log n)$ bits, hence dealing with congestion 
is the main issue.)
\commabsend
Specifically, tight bounds are established in \cite{KKP11} for the time 
and message complexities of the problem of deciding whether a subgraph 
is an MST, and time lower bounds for many other subgraph-decision problems 
(e.g., spanning tree, connectivity) are established in \cite{DHKKNPPW}. 
It is interesting to note that some of these lower bounds imply 
strong unconditional time lower bounds on the hardness of distributed 
approximation for many classical construction problems in the $\cal{CONGEST}$ model.
Decision problems have received recent attention from the asynchrony-discipline too, in the framework of wait-free computing \cite{FRT10}. In this framework, the focus is on task checkability. Wait-free checkable tasks have been characterized in term of covering spaces, a fundamental tool in algebraic topology. 



The theory of \emph{proof-labeling schemes} \cite{KK07,KKP10} was designed 
to tackle the issue of locally verifying 
(with the aid of a {\em proof}, i.e., a certificate, at each node)
solutions to problems that cannot be decided locally 
(e.g.,``is the given subgraph a spanning tree of the network?'',
or, ``is it an MST?''). 
In fact, the model of proof-labeling schemes has some resemblance 
to our definition of the class $\NLD(O(1))$. Investigations in the framework of 
proof-labeling schemes mostly focus on the minimum size of the certificate 
necessary so that verification can be performed in a single round. 
The notion of proof-labeling schemes also has interesting similarities with 
the notions of {\em local detection} \cite{AKY97}, {\em local checking} 
\cite{APV}, or {\em silent stabilization} \cite{silent}, which were introduced 
in the context of {\em self-stabilization}~\cite{D74}. 

The use of oracles that provide information to nodes was studied  intensively in the context of distributed construction tasks.
For instance, this framework, called {\em local computation with advice}, was studied in \cite{FGIP07} for MST construction and 
in \cite{FGIP07}  for 3-coloring a cycle.

Finally, we note that our notion of NLD seems to be related to the theory of {\em lifts}, e.g., \cite{Linial01}. 


\section{Decision problems and
complexity classes}
\label{s:2}

\subsection{Model of computation}

Let us first recall some basic notions in distributed computing. 
We consider the $\local$ model~\cite{Peleg:book}, which is a standard model 
capturing the essence of locality. In this model, processors are assumed 
to be nodes of a network $G$, provided with arbitrary 
distinct identities, and computation proceeds in fault-free synchronous rounds.
At each round, every processor $v\in V(G)$ exchanges messages of 
unrestricted size with its neighbors in $G$, and performs 
computations on its data. We assume that the number of steps (sequential time) used for the local computation 
made by the node $v$ in some round $r$ is bounded by some function $f_A(H(r,v))$, where $H(r,v)$ denotes the size of the ``history'' seen by node $v$ up to the beginning of round $r$. That is, 
the total  number of bits encoded in the input and the identity of the node, as well as in the incoming messages from previous rounds. Here, we do not impose any restriction
on the growth rate of $f_A$. We would like to point out, however,  that imposing such restrictions, or alternatively, imposing restrictions on the memory used by a node
for local computation, may lead to interesting connections between the theory of locality and classical computational complexity theory.
To sum up, during the execution 
of a distributed algorithm~$A$, all processors  are woken up simultaneously, 
and, initially, a processor is solely aware of it own identity, 
and  possibly to some local input too. 
Then, in each round $r$, every processor $v$
\\
(1) sends messages to its neighbors, 
\\
(2) receives messages from its neighbors, and 
\\
(3) performs at most $f_A(H(r,v))$ computations. \\
\\
After a number of rounds (that may depend on the network $G$ and may vary 
among the processors, simply because nodes have different identities,
potentially different inputs, and are typically located at non-isomorphic 
positions in the network), 
every processor $v$ terminates and outputs 
some value $\out(v)$. 
Consider an algorithm running 
in a network $G$ with input $\inp$ and identity assignment $\id$. The {\em running time} of a node $v$, denoted $T_v(G,\inp,\id)$, is the maximum of the number of rounds until $v$ outputs. 
The {\em running time} of the algorithm, denoted $T(G,\inp,\id)$, is the maximum of the number of rounds until all processors terminate, i.e., 
$T(G,\inp,\id)=\max\{T_v(G,\inp,\id)\mid v\in V(G)\}$. 
Let $t$ be a non-decreasing function of input configurations $(G,\inp,\id)$. (By non-decreasing, we mean that if $G'$ is an induced  subgraph of $G$ and $\inp'$ and $\id'$ are the restrictions of $\inp$ and $\id$, respectively, to the nodes in $G'$, then  $t(G',\inp',\id')\leq t(G,\inp,\id)$.)
We say that an algorithm $A$ has running time at most $t$, if  $T(G,\inp,\id)\leq t(G,\inp,\id)$, for every $(G,\inp,\id)$.
We shall give special attention to the case that $t$ represents a constant function. Note that in general, given $(G,\inp,\id)$, the nodes may not be aware 
of $t(G,\inp,\id)$. On the other hand,
note that, if $t=t(G,\inp,\id)$ is known, then w.l.o.g. one can always assume that a local algorithm running 
in time at most $t$ operates at each node $v$ in two stages: 
(A) collect all information available in $B_G(v,t)$,
the {\em $t$-neighborhood},
or {\em ball of radius $t$} of $v$ in $G$,
including inputs, identities and adjacencies, and 
(B) compute the output based on this information. 


\subsection{Local decision (\LD)}

We now refine some of the above concepts, in order to formally define 
our objects of interest. 
Obviously, a distributed algorithm that runs on a graph $G$ operates 
separately on each connected component of $G$, and nodes of a component $G'$ 
of $G$ cannot distinguish the underlying graph $G$ from $G'$. 
For this reason, we consider  connected graphs only.

\commful\vspace*{-1.5ex}\commfulend
\begin{definition}
A {\em configuration} is a pair $(G,\inp)$ where $G$ is a connected graph,
and every node $v\in V(G)$ is assigned as its {\em local input}
a binary string $\inp(v)\in \{0,1\}^*$.
\end{definition}
\commful\vspace*{-1.5ex}\commfulend

In some problems, the local input of every node is empty, i.e., 
$\inp(v)=\epsilon$ for every $v\in V(G)$, where $\epsilon$ denotes 
the empty binary string. 
Since an
undecidable collection of configurations remains undecidable in the 
distributed setting too, we consider only
decidable collections of configurations. 
Formally, we define the following. 
\begin{definition}
A {\em distributed language} is a 
decidable
collection $\cL$ of configurations.
\end{definition}

In general, there are several possible ways of representing a configuration 
of a distributed language corresponding to standard distributed computing 
problems. 
Some examples considered in this paper are the following.
\commful
\\
$\leader = \{(G,\inp) \mid \parallel\inp\parallel_1=1\}$ 
consists of all configurations 
such that there exists a unique node with local input~1, 
with all the others having local input~0. 
\\
$\consensus= \{(G,(\inp_1,\inp_2)) \mid \exists u\in V(G), \forall v\in V(G), 
\inp_2(v)=\inp_1(u)\}$ consists of all configurations such that all nodes 
agree on the value proposed by some node. 
\\
$\coloring = \{(G,\inp) \mid \forall v\in V(G), \forall w\in N(v), \inp(v)
\neq \inp(w)\}$ where $N(v)$ denotes the (open) neighborhood of $v$, 
that is, all nodes at distance~1 from $v$. 
\\
$\mis = \{(G,\inp) \mid S=\{v\in V(G)~|~\inp(v)=1\}\;\mbox{forms a MIS}\}$. 
\\
$\spanningtree = \{(G,(\mbox{name},\mbox{head})) \mid T=\{e_v=(v,v^+), 
v\in V(G), \mbox{head}(v)= \mbox{name}(v^+)\}$ is a spanning tree of $G \}$ 
consists of all configurations such that the set $T$ of edges $e_v$ between 
every node $v$ and its neighbor $v^+$ satisfying 
$\mbox{name}(v^+)=\mbox{head}(v)$ forms a spanning tree of $G$.
\\ 
(The language $\mst$, for minimum spanning tree, can be defined similarly).
\commfulend
\commabs
\begin{description}
\item{$\leader$} 
$= \{(G,\inp) \mid \parallel\inp\parallel_1\leq 1\}$ 
consists of all configurations 
such that there exists at most one node with local input~1, 
with all the others having local input~0. 
\item{$\consensus$}
$= \{(G,(\inp_1,\inp_2)) \mid \exists u\in V(G), \forall v\in V(G), 
\inp_2(v)=\inp_1(u)\}$ consists of all configurations such that all nodes 
agree on the value proposed by some node. 
\item{$\coloring$} 
$= \{(G,\inp) \mid \forall v\in V(G), \forall w\in N(v), \inp(v)
\neq \inp(w)\}$ where $N(v)$ denotes the (open) neighborhood of $v$, 
that is, all nodes at distance~1 from $v$. 
\item{$\mis$}
$= \{(G,\inp) \mid S=\{v\in V(G)~|~\inp(v)=1\}\;\mbox{forms a MIS}\}$. 
\item{$\spanningtree$}
$= \{(G,(\mbox{name},\mbox{head})) \mid T=\{e_v=(v,v^+), 
v\in V(G), \mbox{head}(v)= \mbox{name}(v^+)\}$ is a spanning tree of $G \}$ 
consists of all configurations such that the set $T$ of edges $e_v$ between 
every node $v$ and its neighbor $v^+$ satisfying 
$\mbox{name}(v^+)=\mbox{head}(v)$ forms a spanning tree of $G$.
\\ 
(The language $\mst$, for minimum spanning tree, can be defined similarly).
\end{description}
\commabsend

An identity assignment $\id$ for a graph $G$ is an assignment of 
distinct integers
to the nodes of $G$. A node $v\in V(G)$ executing 
a distributed algorithm in a configuration $(G,\inp)$ initially knows only 
its own identity $\id(v)$ and its own input $\inp(v)$, and is unaware of 
the graph $G$. After $t$ rounds, $v$ acquires knowledge only of its 
$t$-neighborhood $B_G(v,t)$.
In each round $r$ of the algorithm $A$, a node may communicate with its neighbors by sending and receiving messages, and may perform
at most  $f_A(H(r,v))$ computations. Eventually, each node $v\in V(G)$ must output  a local output $\out(v)\in\{0,1\}^*$.


Let $\cL$ be a distributed language. We say that a distributed algorithm 
$\cA$ \emph{decides} $\cL$ if and only if for every configuration $(G,\inp)$, 
and for every identity assignment $\id$ for the nodes of $G$, every node 
of $G$ eventually terminates and outputs ``yes'' or ``no'', 
satisfying the following decision rules: 
\begin{smallitemize}
\item 
If $(G,\inp)\in \cL$, then $\out(v)=$``yes'' for every node $v\in V(G)$;
\item 
If $(G,\inp)\notin \cL$, then there exists at least one node $v\in V(G)$ 
such that $\out(v)=$``no''.
\end{smallitemize}
We are now ready to define one of our main subjects of interest, 
the class $\LD(t)$, for \emph{local decision}.  

\commful\vspace*{-1.5ex}\commfulend
\begin{definition}
Let $t$ be a non-decreasing  function of triplets $(G,\inp,\id)$. Define $\LD(t)$ as the class of all distributed languages that can be decided 
by a  local distributed algorithm that runs in number of rounds at most $t$.
\end{definition}
\commful\vspace*{-1.5ex}\commfulend

For instance, $\coloring\in\LD(1)$ and $\mis\in\LD(1)$.
On the other hand, it is not hard to see that languages such as 
$\leader$, $\consensus$, and $\spanningtree$ are not in $\LD(t)$, for any $t=o(n)$. 
In what follows,  we define $\LD(O(t))=\cup_{c>1} \LD(c\cdot t)$.

\subsection{Non-deterministic local decision (\NLD)}
\label{subsec:NLD}

A distributed {\em verification} algorithm is a distributed algorithm $A$ 
that gets as input, in addition to a configuration $(G,\inp)$, a global 
{\em certificate vector} $\certif$, i.e., every node $v$ of a graph $G$ 
gets as input a binary string $\inp(v)\in\{0,1\}^*$, and a certificate 
$\certif(v)\in\{0,1\}^*$. A verification algorithm $A$ verifies $\cL$ 
if and only if for every configuration $(G,\inp)$, the following hold: 
\begin{itemize}
\item 
If $(G,\inp)\in \cL$, then there exists a certificate $\certif$ 
such that for every 
id-assignment $\id$, algorithm $A$ applied on $(G,\inp)$ with certificate $\certif$ and id-assignment $\id$ outputs  $\out(v)=$``yes'' for all $v\in V(G)$; 
\item 
If $(G,\inp)\notin \cL$, then for every certificate $\certif$ and for every 
id-assignment $\id$,  algorithm $A$ applied on $(G,\inp)$ with certificate $\certif$ and id-assignment $\id$ outputs $\out(v)=$``no'' for at least one node
 $v\in V(G)$.
\end{itemize}
One motivation for  studying the nondeterministic verification framework 
comes from settings in which one must perform local verifications repeatedly. 
In such cases, one can afford to have a relatively ``wasteful'' 
preliminary step in which a certificate is computed for each node. 
Using these certificates,  local verifications can then be performed 
very fast. See \cite{KK07,KKP10} for more details regarding such applications.
Indeed, the definition of a verification algorithm finds similarities 
\commful
with the  notion of {\em proof-labeling schemes}\footnote{
Informally,  in a proof-labeling scheme, the construction
of a  ``good'' certificate~$\certif$ for a configuration $(G,\inp)\in \cL$ 
may depend also on the given id-assignment. 
Since the question of whether a configuration $(G,\inp)$ belongs 
to a language $\cL$ is independent from the particular id-assignment, 
we prefer to let the ``good'' certificate $\certif$ depend only on 
the configuration. In other words,  as defined above, a verification 
algorithm operating on a configuration $(G,\inp)\in \cL$ and a ``good'' 
certificate $\certif$ must say ``yes'' at every node regardless of 
the id-assignment.}
discussed  in \cite{KK07,KKP10}. 
\commfulend
\commabs
with the  notion of {\em proof-labeling schemes}
discussed  in \cite{KK07,KKP10}. 
Informally,  in a proof-labeling scheme, the construction
of a  ``good'' certificate~$\certif$ for a configuration $(G,\inp)\in \cL$ 
may depend also on the given id-assignment. 
Since the question of whether a configuration $(G,\inp)$ belongs 
to a language $\cL$ is independent from the particular id-assignment, 
we prefer to let the ``good'' certificate $\certif$ depend only on 
the configuration. In other words,  as defined above, a verification 
algorithm operating on a configuration $(G,\inp)\in \cL$ and a ``good'' 
certificate $\certif$ must say ``yes'' at every node regardless of 
the id-assignment.
\par 
\commabsend
We now define the class $\NLD(t)$, for \emph{nondeterministic local decision}.
\commabs 
(our terminology is by direct analogy to the class NP in sequential 
computational complexity). 
\commabsend

\commful\vspace*{-1.5ex}\commfulend
\begin{definition}
Let $t$ be a non-decreasing  function of triplets $(G,\inp,\id)$. Define $\NLD(t)$ as the class of all distributed languages that can be verified in at most $t$ communication rounds.
\end{definition}
\commful\vspace*{-1.5ex}\commfulend

\commful\parskip-0.1in\commfulend

\subsection{Bounded-error probabilistic local decision (\BPLD)}

A {\em randomized} distributed algorithm is a distributed algorithm $A$ that  enables every node $v$, at any round $r$ during the execution,  to toss a number of random 
bits obtaining a string $r(v)\in\{0,1\}^*$. Clearly, this  number 
cannot exceed $f_A(H(r,v))$, the bound on the number of computational steps used by node $v$ at round $r$. Note however, that $H(r,v)$ may now also depend
on the random bits produced by other nodes in previous rounds.
 For $p,q\in(0,1]$, we say that 
a randomized distributed algorithm $\cA$ is a {\em $(p,q)$-decider} for $\cL$,
or, that it decides $\cL$ with ``yes'' success  probability $p$ and 
``no'' success  probability $q$, if and only if for every configuration 
$(G,\inp)$, and for every identity assignment $\id$ for the nodes of $G$, 
every node of $G$ eventually terminates and outputs ``yes'' or ``no'',
and the following properties are satisfied: 
\begin{smallitemize}
\item 
If $(G,\inp)\in \cL$, then 
$\Pr[\out(v)=\mbox{``yes'' for every node}~v\in V(G)] \ge p$,
\item 
If $(G,\inp)\notin \cL$, then 
$\Pr[\out(v)=\mbox{``no'' for at least one node}~v\in V(G)] \ge q$,
\end{smallitemize}
where the probabilities
\commabs
in the above definition 
\commabsend
are taken over all possible coin tosses performed by nodes.
We define the class $\BPLD(t,p,q)$, for 
``Bounded-error Probabilistic Local Decision'', as follows. 

\commful\vspace*{-1.5ex}\commfulend
\begin{definition}
For $p,q\in(0,1]$ and a function $t$, $\BPLD(t,p,q)$ is the class of all distributed languages that 
have a local randomized distributed $(p,q)$-decider running in time $t$.
\commabs
(i.e., can be decided in time $t$ by a local randomized distributed algorithm 
with ``yes'' success  probability $p$ and ``no'' success probability $q$).
\commabsend
\end{definition}
\commful\vspace*{-1.5ex}\commfulend

\section{A sharp threshold for randomization}

Consider some graph $G$, and a subset $U$ of the nodes of $G$, i.e., 
$U\subseteq V(G)$. Let $G[U]$ denote the vertex-induced subgraph of $G$ 
defined by the nodes in $U$. Given a configuration $(G,\inp)$, let $\inp[U]$ 
denote the input  $\inp$ restricted to the nodes in $U$. For simplicity of 
presentation, if $H$ is a subgraph of $G$, we denote  $\inp[V(H)]$ by  
$\inp[H]$. A {\em prefix} of a configuration $(G,\inp)$ is a configuration 
$(G[U],\inp[U])$, where $U\subseteq V(G)$  (note that in particular, 
$G[U]$ is connected). We say that a language $\cL$ is {\em hereditary} 
if every prefix of every configuration $(G,\inp) \in \cL$ is also in $\cL$. 
 $\coloring$ and $\leader$ are clearly  hereditary languages. 
As another example of an hereditary language, consider a family $\cG$  
of hereditary graphs, i.e., that is closed under vertex deletion; 
then the language $\{(G,\epsilon)\mid G\in \cG\}$ is hereditary. 
Examples of hereditary graph families are  planar graphs, interval graphs, 
forests, chordal graphs, cographs, perfect graphs, etc.

Theorem \ref{theorem-RLD-closed} below asserts that, for hereditary languages, 
randomization does not help if one imposes that
$p^2+q>1$,
\commabs
i.e, the "no" success  probability distribution is at least as large as 
one minus the square of the "yes" success  probability. 
\commabsend
Somewhat more formally, we prove that for hereditary languages, we have
$\bigcup_{p^2+q> 1} \BPLD(t,p,q)=\LD(O(t))$. This complements the fact that for $p^2+q\leq 1$, we have $\LD(t)\subsetneq  \BPLD(t,p,q)$, for any $t= o(n)$.

Recall that \cite{NS93} investigates the question of whether randomization 
helps for constructing in constant time a solution for a problem in LCL$\subsetneq LD(O(1))$. 
We stress that the technique used in \cite{NS93} for tackling this question 
relies heavily on the definition of 
LCL, specifically, that only graphs of constant degree and of constant input 
size are considered.  
Hence it is not clear whether the technique of \cite{NS93} can be useful 
for our purposes, as we impose no such assumptions on the degrees 
or input sizes. Also, although it seems at first glance, that Lov‡sz local lemma might have been helpful here,  we could not effectively apply it in our proof. 
Instead, we use a completely different approach.

\commful\vspace*{-1.5ex}\commfulend
\begin{theorem}\label{theorem-RLD-closed}
Let $\cL$ be an hereditary language and let $t$ be a  function. If $\cL \in \BPLD(t,p,q)$ 
for constants $p,q\in(0,1]$ such that  $p^2+q> 1$, then $\cL\in \LD(O(t))$. 
\end{theorem}
\commful\vspace*{-1.5ex}\commfulend

\begin{proof}
Let us start with some definitions.  Let  $\cL$ be a language in $\BPLD(t,p,q)$ 
where $p,q\in(0,1]$ and $p^2+q> 1$, and $t$ is some function.
Let $A$ be a randomized algorithm deciding $\cL$, 
with "yes" success  probability $p$, 
and "no" success  probability $q$, whose running time is at most $t(G,\inp,\id)$, for every configuration $(G,\inp)$ with identity assignment $\id$.
Fix a configuration $(G,\inp)$,  and 
an id-assignment $\id$ for the nodes of $V(G)$.
The distance $\dist_G(u,v)$ between two nodes of $G$ is the minimum 
number of edges in a path connecting $u$ and $v$ in $G$.  
The distance between two subsets $U_1,U_2\subseteq V$ is defined as
\commful
$\dist_G(U_1,U_2)=\min\{\dist_G(u,v)\mid u\in U_1, v\in U_2\}.$
\commfulend
\commabs
$$\dist_G(U_1,U_2) ~=~ \min\{\dist_G(u,v)\mid u\in U_1, v\in U_2\}.$$
\commabsend

For  a set $U\subseteq V$, 
let $\cE(G,\inp,\id,U)$ denote the event that when running $A$ on $(G,\inp)$ 
with id-assignment $\id$, all nodes in $U$ output ``yes''. 
Let $v\in V(G)$. The running time of $A$ at $v$ may depend on the coin tosses made by the nodes. 
Let $t_v=t_v(G,\inp,\id)$ denote the maximal running time of $v$ over all possible coin tosses. Note that $t_v\leq t(G,\inp,\id)$ (we do not assume that neither $t$ or $t_v$ are known to $v$). 

The {\em radius} of a node $v$, denoted $r_v$, is the maximum  value $t_u$ such that there exists a node $u$, where $v\in B_G(u,t_u)$.
(Observe that the radius of a node is at most $t$.) The radius of a set of nodes $S$ is $r_S:=\max\{r_v\mid v\in S\}$.
In what follows, fix a constant $\delta$ such that $0<\delta<p^2+q-1$, and define 
$\lambda= 11\left\lceil{\log p}/{\log (1-\delta)}\right\rceil$.


A {\em splitter}  of $(G,\inp,\id)$ is a triplet $(S,U_1,U_2)$ 
of pairwise disjoint subsets of nodes such that $S\cup U_1\cup U_2=V$, 
$\dist_G(U_{1},U_2)\geq \lambda r_S$.  (Observe that  $r_S$ may depend on the identity assignment and the input, and therefore, 
being a splitter is not just a topological property depending only on $G$).
Given a splitter $(S,U_1,U_2)$ of $(G,\inp,\id)$,
let $G_{k}=G[U_k\cup S]$, and let $\inp_k$ be the input $\inp$ restricted to nodes in $G_k$, for $k=1,2$. 

The following structural claim  does not use the fact that $\cL$ is hereditary.  

\commful\vspace*{-1.5ex}\commfulend
\begin{lemma}\label{lem:closed}
For every configuration $(G,\inp)$ with identity assignment $\id$, 
and every splitter $(S,U_1,U_2)$ of $(G,\inp,\id)$, we have
$$\Big((G_{1},\inp_1)\in \cL  ~~\mbox{\rm and}~~ (G_{2},\inp_2)\in \cL\Big) 
~~~~ \Rightarrow ~~~~ (G,\inp)\in \cL.$$
\end{lemma}
\commful\vspace*{-1.5ex}\commfulend

Let $(G,\inp)$ be a configuration with identity assignment $\id$.
Assume, towards contradiction, that there exists a splitter $(S,U_1,U_2)$ 
of triplet  $(G,\inp,\id)$, such that $(G_{1},\inp_1)\in \cL$ and $(G_{2},\inp_2)\in \cL$, 
yet $(G,\inp)\notin \cL$.
(The fact that $(G_{1},\inp_1)\in \cL$ and $(G_{2},\inp_2)\in \cL$ implies 
that  both $G_1$ and $G_2$ are connected,  however, we note, that for the 
claim to be true, it is not required that $G[U_1]$, $G[U_2]$ or $G[S]$ 
are connected.) 
Let $d=\lambda r_S$.

Given a vertex $u\in S$, we define the {\em level} of $u$ by 
$\ell(u)=\dist_G(U_1,\{u\})$.
For an integer $i\in[1,d]$, let $L_i$ denote the set of nodes in $S$ of level 
$i$. For an integer $i\in(r_S, d-r_S)$, let $S_i=\bigcup_{j=i-r_S}^{i+r_S} L_j$, 
and finally, for a set $J\subseteq (r_S, d-r_S)$ of integers, let 
$S_J=\bigcup_{i\in J} S_i$. 

Define
$$I ~=~ \{i\in(2r_S, d-2r_S)\mid \Pr[\cE(G,\inp,\id,S_i)] < 1-\delta\} .$$

\commful\vspace*{-1.5ex}\commfulend
\begin{claim}\label{claim:i}
There exists $i\in(2r_S, d-2r_S)$ such that $i\notin I$. 
\end{claim}
\commful\vspace*{-1.5ex}\commfulend

%
%

\begin{proof}
For proving Claim \ref{claim:i}, we  upper bound the size of $ I$ by $d-4r_S-2$. 
This is done 
by covering the integers in $(2r_S,d-2r_S)$ by at most $4r_S+1$ sets, 
such that each one is $(4r_S+1)$-independent, that is, for every two integers 
in the same set, they are at least $4r_S+1$ apart. 
Specifically, for $s\in[1,4r_S+1]$ and $m(S) = \lceil (d-8r_S)/(4r_S+1)\rceil$, 
we define $J_s=\{s+2r_S+j(4r_S+1)\mid j\in [0,m(S)] \}$. Observe that, as desired, 
$(2r_S,d-2r_S)\subset\bigcup_{s\in [1,4r_S+1]}  J_s$, and for each $s\in[1,4r_S+1]$, 
$J_{s}$ is $(4r_S+1)$-independent. In what follows, fix $s \in[1,4r_S+1]$ 
and let $J=J_s$. Since $(G_{1},\inp_1)\in \cL$, we know that, 
$$\Pr[\cE(G_1,\inp_1,\id,S_{J\cap  I})] ~\geq~ p~.$$

Observe that for $i\in(2r_S, d-2r_S)$, $t_v\leq r_v\leq r_S$, and hence, the $t_v$-neighborhood in $G$ of every node 
 $v\in S_i$ is contained in $S\subseteq G_{1}$, i.e., $B_G(v,t_v)\subseteq G_1$. 
It therefore follows that: 
\begin{equation}
\label{eq:1}
\Pr[\cE(G,\inp,\id,S_{J\cap  I})]=\Pr[\cE(G_1,\inp_1,\id,S_{J\cap  I})]\geq p~.
\end{equation}
Consider two integers $a$ and $b$ in $J$. We know that $|a-b|\geq 4r_S+1$. 
Hence, the distance in $G$ between any two nodes $u \in S_a$ and $ v \in S_b$ 
is at least $2r_S+1$. Thus,  the events $\cE(G,\inp,\id,S_a)$ and 
$\cE(G,\inp,\id,S_b)$ are independent. 
It follows by the definition of $ I$, that 
\begin{equation}
\label{eq:2}
\Pr[\cE(G,\inp,\id,S_{J\cap  I})] < (1-\delta)^{|J\cap  I|}
\end{equation}
By (\ref{eq:1}) and (\ref{eq:2}), we have that $p<(1-\delta)^{|J\cap  I|}$ 
and thus $|J\cap  I| < \log p /\log (1-\delta)$. Since $(2r_S, d-2r_S)$ can be covered 
by  the sets $J_s$, $s=1,\dots,4r_S+1$, each of which is $(4r_S+1)$-independent, 
we get that
$$|I| ~=~ \sum_{s=1}^{4r_S+1} |J_s\cap  I| ~<~ (4r_S+1)(\log p /\log (1-r)) ~.$$
Combining this bound with the fact that $d=\lambda r_S$, we get that $d-4r_S-1>|I|$. It follows by the 
pigeonhole principle that there exists some $i\in (2r_S,d-2r_S) $ such that 
$i\notin  I$, as desired. This completes the proof of Claim \ref{claim:i}.
\end{proof}

Fix $i\in (2r_S,d-2r_S) $ such that $i\notin  I$, and let 
$\cF=\cE(G,\inp,\id,S_i)$. By definition, 
\begin{equation}\label{eq:300}
\Pr[\overline{\cF}] \leq \delta<  p^2+q-1. 
\end{equation}
Let $H_1$ denote the subgraph of $G$ induced by the nodes in 
$(\bigcup_{j=1}^{i-r_S-1} L_j) \cup U_1$. We similarly define $H_2$ as 
the subgraph of $G$ induced by the nodes in $(\bigcup_{j>i+r_S} L_j) \cup U_2$. 
Note that $S_i \cup V(H_1) \cup V(H_2)=V$, and for any two nodes $u\in V(H_1)$ and $v\in V(H_2)$, we have $d_G(u,v)>2 r_S$. It follows that, for $k=1,2$, the $t_u$-neighborhood 
in $G$ of each node $u\in V(H_k)$ equals  the $t_u$-neighborhood 
in $G_k$ of $u$, that is, $B_G(u,t_u)\subseteq G_k$. (To see why, consider, for example, the case $k=2$. Given $u\in V(H_2)$, it is sufficient to show that
$\nexists v\in V(H_1)$, such that $v\in B_G(u,t_u)$. Indeed, if such a vertex $v$ exists then $d_G(u,v)>2 r_S$, and hence $t_u>2 r_S$. 
Since  there must exists a vertex $w\in S_i$ such that $w\in B(u,t_u)$, we get that $r_w>2 r_S$, in contradiction to the fact that $w\in S$.)
Thus, for $k=1,2$, since $(G_{i},\inp_i)\in\cL$, we get
$\Pr[\cE(G,\inp,\id,V(H_i))] ~=~ \Pr[\cE(G_{i},\inp_i,\id,V(H_i))] ~\geq~ p~$.

Let $\cF'=\cE(G,\inp,\id,V(H_1)\cup V(H_2))$. As the events 
$\cE(G,\inp,\id,V(H_1))$ and $\cE(G,\inp,\id,V(H_2))$ are independent, 
it follows that $\Pr[\cF'] > p^2$, that is 
\begin{equation}
\label{eq:400}
\Pr[\overline{\cF'}] \leq 1- p^2
\end{equation}
By Eqs. (\ref{eq:300}) and (\ref{eq:400}), and using union bound, it follows 
that $\Pr[\overline{\cF}\vee\overline{\cF'}]< q$. Thus 
$$\Pr[\cE(G,\inp,\id,V(G))] ~=~ \Pr[\cE(G,\inp,\id,S_i  \cup V(H_{1}) 
\cup V(H_2))] ~=~ \Pr[\cF \wedge\cF']> 1-q~. $$
This is in contradiction to the assumption that $(G,\inp)\notin \cL$. 
This concludes the proof of Lemma~\ref{lem:closed}. \qed

Our goal now is to show that $\cL\in \LD(O(t))$ by proving the existence of a deterministic local 
algorithm $D$ that  runs in time $O(t)$ and recognizes $\cL$. (No attempt is made here to minimize the constant factor hidden in the $O(t)$ notation.) 
 Recall that both $t=t(G,\inp,\id)$ and $t_v=t_v(G,\inp,\id)$ may not be known to $v$.
Nevertheless, by inspecting the balls $B_G(v,2^i)$
for increasing $i=1,2,\cdots$, each node $v$ can compute an upper bound on $t_v$ as given by the following claim.

\begin{claim}
Fix a a configuration $(G,\inp)$,  an id-assignment $\id$, and a constant $c$.   In $O(t)$ time, each node $v$ can compute a value $t^*_v=t^*_v(c)$ such that (1) $c\cdot t_v\leq t^*_v=O(t)$ and (2) for every $u\in B_G(v,c\cdot t^*_v)$, we have $t_u\leq t^*_v$.
\end{claim}

To establish the claim, observe first that  in $O(t)$ time, each node $v$ can compute a value $t'_v$ satisfying  $t_v\leq t'_v\leq 2 t$.
Indeed, given the ball $B_G(v,2^i)$, for some integer $i$, and using the upper bound on number of (sequential) local computations, node $v$ can simulate
all its possible executions up to round $r=2^i$. The desired value $t'_v$ is the smallest $r=2^i$ for 
which all executions of $v$ up to round $r$ conclude with an output at $v$. 
Once $t'_v$ is computed, node $v$ aims at computing $t_v^*$. For this purpose, it starts again to inspect the balls
$B_G(v,2^i)$
for increasing $i=1,2,\cdots$,  to obtain $t'_u$ from each $u\in B_G(v,2^i)$. (For this purpose, it may need to wait until $u$ computes $t'_u$, but 
this delays the whole computation by at most $O(t)$ time.) Now, node $v$ outputs $t^*_v=2^i$ for the smallest $i$ satisfying 
(1) $c\cdot t'_v\leq 2^i$ and (2) for every $u\in B_G(v,c\cdot 2^i)$, we have $t'_u\leq t^*_v$.
It is easy to see that for this $i$, we have $2^i=O(t)$, hence $t^*_v=O(t)$.

Given a configuration $(G,\inp)$, 
and an id-assignment $\id$, Algorithm $D$, applied at a node $u$ first calculates $t^*_u=t^*_u(6\lambda)$, and then
outputs ``yes'' if and only if the $2\lambda   t^*_u$-neighborhood of $u$ in $(G,\inp)$ 
belongs to $\cL$. That is, 
$$\out(u) ~=~ \mbox{``yes''}  ~~~\iff~~~ 
(B_G(u, 2\lambda   t^*_u),\inp[B_G(u, 2\lambda  t^*_u)]) ~\in~ \cL~.$$

Obviously,  Algorithm $D$ is a deterministic algorithm that runs in time $O(t)$. 
We claim that Algorithm $D$ decides $\cL$. Indeed, since $\cL$ is hereditary, 
if $(G,\inp)\in \cL$,  then every prefix of $(G,\inp)$ is also in $\cL$, 
and thus, every node $u$ outputs $\out(u)=$``yes". 
Now consider the case where $(G,\inp)\notin \cL$, and assume by contradiction 
that by applying $D$ on 
$(G,\inp)$ with id-assignment $\id$, every node $u$ outputs $\out(u)=$``yes''. 
Let $U\subseteq V(G)$ be maximal by inclusion, such that $G[U]$ is connected 
and  $(G[U],\inp[U])\in\cL$. 
Obviously, $U$ is not empty, as $(B_G(u, 2\lambda  t^*_v),\inp[B_G(u, 2\lambda  t^*_v)])\in \cL$ for every 
node $u$. On the other hand, 
we have $|U|<|V(G)|$, because $(G,\inp)\notin \cL$.

Let $u\in  U$ be a node with maximal  $t_u$ such that $B_G(u,2 t_u)$ contains a node outside $U$.
Define $G'$ as the subgraph of $G$ induced by  $U \cup V(B_G(u,2t_u))$. 
Observe that $G'$ is connected and that $G'$ strictly contains $U$.
Towards contradiction, our goal is to show that $(G',\inp[G'])\in \cL$.


Let $H$ denote the graph which is maximal by inclusion such that 
$H$ is connected and  
$$B_G(u,2t_u) ~\subset~ H ~\subseteq~ B_G(u,2t_u)\cup(U\cap B_G(u,2\lambda  t^*_u))~.$$
Let $W^1,W^2,\cdots, W^{\ell}$ be the $\ell$ connected components of 
$G[U]\setminus B_G(u,2t_u)$, ordered arbitrarily. Let $W^0$ be the empty graph, 
and for $k=0,1,2,\cdots, \ell$, define the graph 
$Z^k ~=~ H\cup W^0\cup W^1\cup W^2\cup\cdots\cup W^k$. 
Observe that $Z^k$ is connected for each $k=0,1,2,\cdots, \ell$. 
We prove by induction on $k$ that $(Z^k,\inp[Z^k])\in \cL$ for every 
$k=0,1,2,\cdots, \ell$. 
This will establish the contradiction since $Z^{\ell}=G'$.
For the basis of the induction, the case $k=0$, we need to show that 
$(H,\inp[H])\in \cL$. However, this is immediate by the facts that $H$ 
is a connected subgraph of $B_G(u,2\lambda t^*_u)$, the configuration 
$(B_G(u, 2\lambda t^*_u),\inp[B_G(u,2\lambda t^*_u)])\in\cL$, and $\cL$ is hereditary. 
Assume now that we have $(Z^k,\inp[Z^k])\in \cL$ for $0\leq k<\ell$, 
and consider  the graph $Z^{k+1}=Z^k\cup W^{k+1}$.
Define the sets of nodes
$$S ~=~ V(Z^k)\cap V(W^{k+1}),~~~ U_1 ~=~ V(Z^k)\setminus S,~~~{\mbox{and}}~~~ 
U_2 ~=~ V(W^{k+1})\setminus S~.$$
A crucial observation is that $(S,U_1,U_2)$ is a splitter of $Z^{k+1}$. 
This follows from the following arguments. Let us first show that $r_S\leq t^*_u$. 
By definition, we have $t_v\leq t^*_u$, for every $v\in  B_G(u, 6\lambda t^*_u)$.
Hence, in order to bound the radius of $S$ (in $Z^{k+1}$) by $t^*_u$ it is sufficient to prove that there is no node $w\in U\setminus B_G(u, 6\lambda t^*_u)$ such that 
$B_G(w, t_w)\cap S\neq \emptyset$. Indeed, if such a node $w$ exists then $t_w>4\lambda t^*_u$ and hence $B_G(w, 2t_w)$ contains a node outside $U$, in contradiction to the choice of $u$. It follows that $r_S\leq t^*_u$. 

We now claim that $\dist_{Z^{k+1}}(U_1,U_2)\geq \lambda t^*_u$. 
Consider a simple directed path $P$ in $Z^{k+1}$ going from a node 
$x\in U_1$ to a node $y\in U_2$.
Since $x\notin V(W^{k+1})$ and $y\in V(W^{k+1})$, we get that $P$ must pass 
through a vertex in $ B_G(u,2t_u)$. Let $z$ be the last vertex in $P$ such that 
$z\in B_G(u,2t_u)$, and  consider  the directed subpath $P_{[z,y]}$ of $P$ going 
from $z$ to $y$.  Now, let $P'=P_{[z,y]}\setminus\{z\}$.
The first $d'=\min\{(2\lambda -2) t^*_u,|P'|\}$ vertices in the directed subpath $P'$ must belong 
to $V(H)\subseteq V(Z^k)$. In addition, observe that all nodes in $P'$ 
must be in $V(W^{k+1})$. It follows that the first $d'$ nodes of $P'$ 
are in $S$. Since $y\notin S$, we get that $|P'|\geq d'=(2\lambda -2) t^*_u$, and thus  
$|P|>\lambda  t^*_u$. Consequently, $\dist_{Z^{k+1}}(U_1,U_2)\geq \lambda  t^*_u$, as desired.
This completes the proof that $(S,U_1,U_2)$ is a splitter of $Z^{k+1}$. 

Now, by the induction hypothesis, we have $(G_1,\inp[G_1])\in \cL$, 
because $G_1=G[U_1\cup S]=Z^k$. In addition, 
we have $(G_2,\inp[G_2])\in \cL$, because $G_2=G[U_2\cup S]=W^{k+1}$, 
and $W^{k+1}$ is a prefix of $G[U]$.
We can now apply Lemma~\ref{lem:closed} and conclude that 
$(Z^{k+1},\inp[Z^{k+1}])\in \cL$. This concludes the induction proof. 
The theorem follows.
\end{proof}

Let ${\tt Planar}=\{(G,\epsilon):\mbox{$G$ is planar}\}$, ${\tt Interval}=\{(G,\epsilon):\mbox{$G$ is an interval graph}\}$  and 
${\tt CycleFree}=\{(G,\epsilon):\mbox{$G$ has no cycle}\}$. 
One can easily check that neither of these three languages 
is in $\LD(t)$, for any $t=o(n)$. 
Hence, Theorem~\ref{theorem-RLD-closed} yields the following.

\commful\vspace*{-1.5ex}\commfulend
\begin{corollary}
Let $p,q\in(0,1]$ such that $p^2+q> 1$, then $\leader$, ${\tt Planar}$, ${\tt Interval}$ and
${\tt CycleFree}$ are not in  $\BPLD(t,p,q)$, for any $t=o(n)$. 
\end{corollary}
\commful\vspace*{-1.5ex}\commfulend

\section{Nondeterminism and complete problems}

\subsection{Separation results}

\commful\parskip0.0in\commfulend

Our first separation result indicates that non-determinism helps for local decision. Indeed, we show that there exists a language, specifically, $\tree=\{(G,\epsilon) \mid G\;\mbox{is a tree}\}$, which belongs to $\NLD(1)$ but not to $\LD(t)$, for any $t=o(n)$. The proof follows by rather standard arguments.

\commful\vspace*{-1.5ex}\commfulend
\begin{theorem}\label{separate-LD}
$\LD(t)\subsetneq \NLD(t)$, for any $t=o(n)$.
\end{theorem}
\commful\vspace*{-1.5ex}\commfulend

\def\APPENDIXD{
To establish the theorem it is sufficient to show that there exists 
a language $\cL$ such that $\cL\notin \LD(o(n))$ and $\cL\in \NLD(1)$. 
Let $\tree=\{(G,\epsilon) \mid G\;\mbox{is a tree}\}$. 
We have  $\tree\notin LD(o(n))$. To see why, consider a cycle $C$ 
with nodes labeled consecutively from~1 to~$4n$, and the path $P_1$ 
(resp., $P_2$) with nodes labeled consecutively $1,\dots,4n$ 
(resp., $2n+1,\dots,4n,1,\dots,2n$), from one extremity to the other. 
For any algorithm $A$ deciding $\tree$,  all nodes $n+1,\dots,3n$ 
output ``yes''  in configuration $(P_1,\epsilon)$ for any 
identity assignment for the nodes in $P_1$, while all nodes 
$3n+1,\dots,4n,1,\dots,n$ output ``yes''  in configuration $(P_2,\epsilon)$ 
for any identity assignment or the nodes in $P_2$.  
Thus if $A$ is local, then all nodes output ``yes'' in configuration 
$(C,\epsilon)$, a contradiction. 
In contrast, we next show that $\tree\in \NLD$. 
The (nondeterministic) local algorithm $A$ verifying $\tree$ 
operates as follows. Given a configuration $(G,\epsilon)$, 
the certificate given at node $v$ is $\certif(v)=\dist_G(v,r)$ 
where $r\in V(G)$ is an arbitrary fixed node. The verification procedure 
is then as follows. At each node $v$, $A$ inspects every neighbor 
(with its certificates), and verifies the following: 
\begin{itemize}
\item 
$\certif(v)$ is a non-negative integer,
\item 
if $\certif(v)=0$, then $\certif(w)=1$ for every neighbor $w$ of $v$, and 
\item 
if $\certif(v)> 0$, then there exists a neighbor $w$ of $v$ such that 
$\certif(w)=\certif(v)-1$, and, for all other neighbors $w'$ of $v$, 
we have $\certif(w')=\certif(v)+1$.
\end{itemize}
If $G$ is a tree, then applying Algorithm $A$ on $G$ with the certificate 
yields the answer ``yes'' at all nodes regardless of the given id-assignment. 
On the other hand, if $G$ is not a tree, then we claim that 
for every certificate, and every id-assignment $\id$, Algorithm $A$ outputs ``no'' at some node. 
Indeed, consider some certificate $\certif$ given to the nodes of $G$, 
and let $C$ be a simple cycle in $G$. Assume, for the sake of contradiction,
that all nodes in $C$ output ``yes''. In this case, each node in $C$ has at least one neighbor in $C$ with a larger certificate. This creates an infinite sequence of strictly increasing certificates, in contradiction with the finiteness of~$C$. 
} 

\commabs
\begin{proof}
\APPENDIXD
\end{proof}
\commabsend

\commful\vspace*{-1.5ex}\commfulend
\begin{theorem}\label{separate-NLD}
There exists a language $\cL$ such that $\cL\notin \NLD(t)$, for any $t=o(n)$.
\end{theorem}
\commful\vspace*{-1.5ex}\commfulend

\def\APPENDIXE{
Let $\inpsize=\{(G,\inp) \mid \forall v\in V(G),\; \inp(v)=|V(G)|\}$. 
We show that $\inpsize\notin \NLD(t)$, for any $t=o(n)$. Assume, for the sake of contradiction, that there exists a local nondeterministic algorithm $A$ deciding $\inpsize$. Let $t<n/4$ be the running time of $A$. Consider the cycle $C$ with $2t+1$ nodes $u_1,u_2,\cdots, u_{2t+1}$, enumerated clockwise. Assume that the input  at each node $u_i$ of $C$ satisfies $\inp(u_i)=2t+1$. Then, there exists a certificate $\certif$ such that, for any identity assignment $\id$, algorithm $A$ outputs ``yes'' at each node of $C$. Now, consider the configuration $(C',\inp')$ where the cycle $C'$ has $4t+2$ nodes, and for each node $v_i$ of $C'$,  $\inp'(v_i)=2t+1$. We have $(C',\inp')\notin \inpsize$. To fool Algorithm $A$, we enumerate the nodes in $C'$ clockwise, i.e., $C=(v_1,v_2,\cdots, v_{4t+2})$. We then define the certificate $\certif'$ as follows: 
$$\certif'(v_i) ~=~ \certif'(v_{i+2t+1}) ~=~ \certif(u_i)
~~\mbox{for $i=1,2,\cdots 2t+1$}~â€ .$$

Fix an id-assignment $\id'$ for the nodes in $V(C')$, and fix $i\in\{1,2,\cdots 2t+1\}$.
There exists an  id-assignment $\id_1$ for the nodes in $V(C)$, such that 
the output of $A$ at node $v_i$  in $(C',\inp')$ with certificate $\certif'$ and id-assignment $\id'$ is identical to the output of $A$ at node $u_i$ in $(C,\inp)$
with certificate $\certif$ and id-assignment $\id_1$. Similarly, there exists an  id-assignment $\id_2$ for the nodes in $V(C)$ such that 
the output of $A$ at node $v_{i+2t+1}$  in $(C',\inp')$ with certificate $\certif'$ and  id-assignment $\id'$ is identical to the output of $A$ at node $u_i$ in $(C,\inp)$
with with certificate $\certif$  and id-assignment $\id_2$. 
Thus, Algorithm $A$ at both $v_i$ and  $v_{i+2t+1}$  outputs ``yes''  in $(C',\inp')$ with certificate $\certif'$ and id-assignment $\id'$. Hence,
since $i$ was arbitrary,  all nodes  output ``yes'' for this configuration, certificate and id-assignment, contradicting the fact that  $(C',\inp')\notin \inpsize$. 
} 

\commabs
\begin{proof}
\APPENDIXE
\end{proof}
\commabsend

\commful\parskip-0.1in\commfulend

 For $p,q\in(0,1]$ and a function $t$, let us define $\BPNLD(t,p,q)$ as the class of all distributed languages that 
have a local randomized non-deterministic distributed $(p,q)$-decider running in time $t$.  
\commful\vspace*{-1.5ex}\commfulend
\begin{theorem}\label{theo:all}
Let $p,q\in(0,1]$  such that $p^2+q\leq 1$. For every language $\cL$, we have $\cL\in \BPNLD(1,p,q)$.
\end{theorem}
\commful\vspace*{-1.5ex}\commfulend

\def\APPENDIXC{
Let $\cL$ be a language. The certificate of a  configuration $(G,\inp)\in\cL$ is a map of $G$, with nodes labeled with distinct integers in $\{1,...,n\}$, where $n=|V(G)|$, together with the inputs of all nodes in $G$. In addition, every node $v$ receives the label $\lambda(v)$ of the corresponding vertex in the map. Precisely, the certificate at node $v$ is  $\certif(v)=(G',\inp',i)$ where $G'$ is an isomorphic copy of $G$ with nodes labeled from~1 to~$n$, $\inp'$ is an $n$-dimensional vector such that $\inp'[\lambda(u)]=\inp(u)$ for every node $u$, and $i=\lambda(v)$. The verification algorithm involves checking that the configuration $(G',\inp')$ is identical to $(G,\inp)$. This is sufficient because distributed languages are sequentially decidable, hence every node can individually decide whether $(G',\inp')$ belongs to $\cL$ or not, once it has secured the fact that $(G',\inp')$ is the actual configuration. It remains to show that there exists a local randomized non-deterministic distributed $(p,q)$-decider for  verifying  that the configuration $(G',\inp')$ is identical to $(G,\inp)$, and running in time 1.

The non-deterministic $(p,q)$-decider operates as follows. First, 
every node $v$ checks that it has received the input as specified by $\inp'$, i.e., $v$ checks wether $\inp'[\lambda(v)]=\inp(v)$, and outputs ``no'' if this does not hold. Second, each node $v$ communicates with its neighbors to check that (1) they all got the same map $G'$ and the same input vector $\inp'$, and (2) they are labeled the way they should be according to the map $G'$. If some inconsistency is detected by a node, then this node outputs ``no''. 
Finally, consider a node  $v$ that passed the aforementioned  two phases without outputting ``no'' . If $\lambda(v)\neq 1$ then $v$   outputs ``yes'' (with probability 1), and
if  $\lambda(v)= 1$ then $v$  outputs ``yes'' with probability $p$.

We claim that the above implements a non-deterministic distributed $(p,q)$-decider for  verifying  that the configuration $(G',\inp')$ is identical to $(G,\inp)$.
 Indeed, if all nodes pass the two phases without outputting ``no'', then they all agree on the map $G'$ and on the input vector $\inp'$,  and they know that their respective neighborhood fits with what is indicated on the map. 
Hence,  $(G',\inp')$ is a lift of $(G,\inp)$. If follows that $(G',\inp')=(G,\inp)$ if and only if there exists at most one node $v\in G$, whose label satisfies $\lambda(v)= 1$. Consequently,
if $(G',\inp')=(G,\inp)$ then all nodes say ``yes'' with probability at least $p$. On the other hand,  if $(G',\inp')\neq (G,\inp)$ then there are at least two nodes in $G$ whose label is ``1''. These two nodes say ``yes'' with probability  $p^2$, hence, the probability that at least one of them says ``no'' is at least $1-p^2\geq q$.
%
%
This completes the proof of Theorem \ref{theo:all}.  
} 

\commabs
\begin{proof}
\APPENDIXC
\end{proof}
\commabsend

The above theorem guarantees that the following definition is well defined. Let $\BPNLD=\BPNLD(1,p,q)$, for some $p,q\in(0,1]$ such that $p^2+q\leq 1$.
The following follows from Theorems \ref{separate-LD},  \ref{separate-NLD} and \ref{theo:all}.
\begin{corollary}
$\LD(o(n))\subsetneq \NLD(O(1))\subset  \NLD(o(n)) \subsetneq \BPNLD=\mbox{All}$.
\end{corollary}
\subsection{Completeness results}

Let us first define a notion of reduction that fits the class $\LD$. 
For two languages $\cL_1,\cL_2$, we say that $\cL_1$ is 
\emph{locally reducible} to $\cL_2$, denoted by $\cL_1 \preceq  \cL_2$, 
if there exists a constant time local algorithm $A$ such that, for every configuration 
$(G,\inp)$ and every id-assignment $\id$, $A$ produces $\out(v)\in \{0,1\}^*$ 
as output at every node $v\in V(G)$ so that 
$$(G,\inp)\in \cL_1 ~~~\iff~~~ (G,\out)\in\cL_2~.$$
By definition, $\LD(O(t))$ is closed under local reductions, that is, 
for every two languages $\cL_1,\cL_2$ satisfying $\cL_1 \preceq  \cL_2$, 
if $\cL_2\in \LD(O(t))$ then $\cL_1\in \LD(O(t))$. 

We now show that there exists a natural problem,  called 
$\cover$, which is in some sense the ``most difficult'' decision problem; that is, we show that $\cover$ is $\BPNLD$-complete. 
 Language $\cover$ is defined as follows. 
Every node $v$ is given as input an element $\cE(v)$, and a finite collection of sets $\cS(v)$. The union of these inputs is in the language if there exists a node $v$ such that one set in $\cS(v)$ equals the union of all the elements given to the nodes. Formally, we define
$\cover ~=~ \{(G,(\cE,\cS)) \mid \exists v\in V(G),\; \exists S\in \cS(v)\;
\mbox{s.t.}\; 
S=\{ \cE(v) \mid v\in V(G) \}
.$

\commful
The proof of the following Theorem is deferred to 
Appendix \ref{App:theo:hardness2}.
\commfulend

\commful\vspace*{-1.5ex}\commfulend
\begin{theorem}
\label{theo:hardness2}
$\cover$ is $\BPNLD$-complete.
\end{theorem}
\commful\vspace*{-1.5ex}\commfulend
\begin{proof}
The fact that $\cover\in\BPNLD$ follows from Theorem \ref{theo:all}. 
To prove that $\cover$ is  $\BPNLD$-hard, we consider some $\cL\in \BPNLD$
and show that $\cL\preceq\cover$. 
For this purpose, we describe a local distributed algorithm $A$ transforming any configuration for $\cL$ to a configuration for $\cover$ preserving the memberships to these languages. Let $(G,\inp)$ be a configuration for $\cL$ and let $\id$ be an identity assignment. Algorithm $A$ operating at a node $v$ outputs a pair $(\cE(v),\cS(v))$, where $\cE(v)$ is the ``local view'' at $v$ in $(G,\inp)$, i.e., the star subgraph of $G$ consisting of $v$ and its neighbors, together with the inputs of these nodes  and their identities, and $\cS(v)$ is the collection of sets $S$ defined as follows. 
For a binary string $x$, let $|x|$ denote the {\em length} of $x$, i.e., the number of bits in $x$. For every vertex $v$, let $\psi(v)=2^{|\id(v)|+|\inp(v)|}$.
Node $v$ first generates all configurations $(G',\inp')$ where $G'$ is a graph with $k\leq \psi(v)$ vertices, and $\inp'$ is a collection of $k$ input strings of length at most $\psi(v)$, such that $(G',\inp')\in \cL$. For each such configuration $(G',\inp')$, node $v$ generates all possible $\id'$ assignments to  $V(G')$  such that for every node $u\in V(G')$, $|\id(u)|\leq \psi(v)$. Now, for each such pair of a graph $(G',\inp')$ and an $\id'$ assignment, algorithm $A$  associates a set $S\in\cS(v)$ consisting of the $k=|V(G')|$ local views of the nodes of $G'$ in $(G',\inp')$. We show that $(G,\inp)\in\cL \iff A(G,\inp)\in\cover$. 

If $(G,\inp)\in\cL$, then by the construction of Algorithm $A$, there exists a set $S\in \cS(v)$ such that $S$ covers the collection of local views for $(G,\inp)$, i.e., 
$S = \{ \cE(u) \mid u\in G \}$. 
Indeed, the node $v$ maximizing $\psi(v)$ satisfies $\psi(v)\geq \max\{\id(u)\mid u\in V(G)\}\geq n$ and $\psi(v)\geq \max\{\inp(u)\mid u\in V(G)\}$.
 Therefore, that specific node has constructed a set $S$ which contains all local views of the given configuration $(G,\inp)$ and $\id $ assignemnt. Thus $A(G,\inp)\in\cover$.

Now consider the case that $A(G,\inp)\in\cover$. In this case, there exists a node $v$ and a set $S\in\cS(v)$ such that 
$S = \{ \cE(u) \mid u\in G \}$. 
Such a set $S$ is the collection of local views of nodes of some configuration $(G',\inp')\in \cL$ and some $\id'$ assignment. On the other hand, $S$ is also the collection of local views of nodes of the given configuration $(G,\inp)\in \cL$ and  $\id$ assignment. It follows that $(G,\inp)=(G',\inp')\in \cL$.
\end{proof}

%

We now define a natural problem, called $\containment$, which is $\NLD(O(1))$-complete. 
Somewhat surprisingly, the definition of $\containment$ is quite similar to the definition of $\cover$. 
Specifically, as in $\cover$, 
every node $v$ is given as input an element $\cE(v)$, and a finite collection of sets $\cS(v)$. However, in contrast to $\cover$, the union of these inputs is in the  $\containment$ language if there exists a node $v$ such that one set in $\cS(v)$ {\em contains} the union of all the elements given to the nodes. Formally, we define
$\containment ~=~ \{(G,(\cE,\cS)) \mid \exists v\in V(G),\; \exists S\in \cS(v)\;
\mbox{s.t.}\; 
S\supseteq\{ \cE(v) \mid v\in V(G) \}
.$

\commful
The proof of the following Theorem is deferred to 
Appendix \ref{App:theo:hardness2}.
\commfulend

\commful\vspace*{-1.5ex}\commfulend
\begin{theorem}
\label{theo:hardness2}
$\containment $ is $\NLD(O(1))$-complete.
\end{theorem}
\commful\vspace*{-1.5ex}\commfulend
\begin{proof}
We first prove that $\containment $ is  $\NLD(O(1))$-hard. Consider some $\cL\in \NLD(O(1))$; we 
show that $\cL\preceq \containment $. 
For this purpose, we describe a local distributed algorithm $D$ transforming any configuration for $\cL$ to a configuration for $\containment $ preserving the memberships to these languages. 

Let $t = t_{\cL} \geq 0$ be some (constant) integer such that there exists a local 
nondeterministic algorithm $A_{\cL}$ deciding $\cL$ in time at most $t$.
Let $(G,\inp)$ be a configuration for $\cL$ and let $\id$ be an identity assignment. Algorithm $D$ operating at a node $v$ outputs a pair $(\cE(v),\cS(v))$, where $\cE(v)$ is the ``$t$-local view'' at $v$ in $(G,\inp)$, i.e.,  the ball of radius $t$ around $v$, $B_G(v,t)$, together with the inputs of these nodes  and their identities, and $\cS(v)$ is the collection of sets $S$ defined as follows.
For a binary string $x$, let $|x|$ denote the {\em length} of $x$, i.e., the number of bits in $x$. For every vertex $v$, let $\psi(v)=2^{|\id(v)|+|\inp(v)|}$.
Node $v$ first generates all configurations $(G',\inp')$ where $G'$ is a graph with $m\leq \psi(v)$ vertices, and $\inp'$ is a collection of $m$ input strings of length at most $\psi(v)$, such that $(G',\inp')\in \cL$. For each such configuration $(G',\inp')$, node $v$ generates all possible $\id'$ assignments to  $V(G')$  such that for every node $u\in V(G')$, $|\id(u)|\leq \psi(v)$. Now, for each such pair of a graph $(G',\inp')$ and an $\id'$ assignment, algorithm $D$  associates a set $S\in\cS(v)$ consisting of the $m=|V(G')|$ $t$-local views of the nodes of $G'$ in $(G',\inp')$. We show that $(G,\inp)\in\cL \iff D(G,\inp)\in \containment $. 

If $(G,\inp)\in\cL$, then by the construction of Algorithm $D$, there exists a set $S\in \cS(v)$ such that $S$ covers the collection of $t$-local views for $(G,\inp)$, i.e., 
$S = \{ \cE(u) \mid u\in G \}$. 
Indeed, the node $v$ maximizing $\psi(v)$ satisfies $\psi(v)\geq \max\{\id(u)\mid u\in V(G)\}\geq n$ and $\psi(v)\geq \max\{\inp(u)\mid u\in V(G)\}$.
 Therefore, that specific node has constructed a set $S$ 
 that precisely corresponds to $(G,\inp)$ and its given $\id $ assignment; hence, $S$
  contains all corresponding $t$-local views. Thus, $D(G,\inp)\in \containment $.

Now consider the case that $D(G,\inp)\in \containment $. In this case, there exists a node $v$ and a set $S\in\cS(v)$ such that 
$S \supseteq \{ \cE(u) \mid u\in G \}$. 
Such a set $S$ is the collection of $t$-local views of nodes of some configuration $(G',\inp')\in \cL$ and some $\id'$ assignment. 
Since $(G',\inp')\in \cL$, there exists a certificate $\certif'$ for the nodes of $G'$, such that when algorithm $A_{\cL}$ operates on 
 $(G',\inp',\certif')$, all nodes say ``yes''. 
Now, since $S$ contains the $t$-local  views of nodes  $(G,\inp)$, with the corresponding identities, there exists a mapping $\phi: (G,\inp,\id) \rightarrow (G',\inp',\id')$  that preserves inputs and identities. Moreover,  when restricted to a ball of radius $t$ around a vertex $v\in G$,
$\phi$ is actually an isomorphism between this ball and its image.
We assign a certificate $\certif$ to the nodes  of $G$: for each $v\in V(G)$, $\certif(v)=\certif'(\phi(v))$. Now,
 Algorithm $A_{\cL}$ when operating on $(G,\inp,\certif)$ outputs ``yes'' at each node of $G$. By the correctness of $A_{\cL}$, we obtain  $(G,\inp)\in \cL$.

We now show that $\containment\in \NLD(O(1))$.
 For this purpose, 
we design a nondeterministic local algorithm $A$ that decides 
whether a configuration $(G,\inp)$ is in $\containment $. 
Such an algorithm $A$ is designed to operate on $(G,\inp,\certif)$, 
where $\certif$ is a certificate. The configuration $(G,\inp)$ satisfies 
that $\inp(v)=(\cE(v),\cS(v))$. Algorithm $A$ aims at verifying 
whether there exists a node $v^*$ with a  set $S^*\in \cS(v^*)$ 
such that $S^*\supseteq\{ \cE(v) \mid v\in V(G)\}$.

Given a correct instance, i.e., a configuration $(G,\inp)$, we define the certificate $\certif$ as follows.
For each node $v$, the certificate $\certif(v)$ at $v$ consists of several fields, specifically, $\certif(v)=(\certif_c(v),\certif_s(v),\certif_{id}(v),\certif_l(v))$. The {\em candidate configuration} field $\certif_{c}(v)$ is 
a triplet $\certif_{c}(v)=(G',\inp',\id')$, where $(G',\inp')$ is an isomorphic copy $(G',\inp')$ of  $(G,\inp)$ and
$\id'$ is an identity assignment for the nodes of $G'$. The {\em candidate set} field $\certif_{s}(v)$ is
a copy of $S^*$, i.e., $\certif_{s}(v)=S^*$. In addition, 
let $u$ and $u^*$ be the nodes in $(G',\inp')$ corresponding to $v$ and $v^*$, respectively. 
The {\em candidate identity} field $\certif_{id}(v)$ is $\certif_{id}(v)=\id'(u)$, and the {\em candidate leader} field $\certif_{l}(v)$ is $\certif_{l}(v)=\id'(u^*)$.

We describe the operation of Algorithm  $A$ on some triplet $(G,\inp,\certif)$. First, each node $v$ verifies that it agrees with its neighbors on the candidate configuration and candidate set fields in their certificates. That way, if all nodes say ``yes'' then we know that all nodes hold the same 
candidate configuration which is some triplet $(G',\inp',\id')$, and the same candidate set $S'$. Second, each node verifies  that  $\cE(v)\in S'$. Also, 
each node checks that it agrees with its neighbors on the candidate leader field in their certificates. I.e, that there exists some integer  $k$
such that for all nodes $v$ we have  $\certif_l(v)=k$. Each node $v$ checks that there exists a node $u^*\in V(G')$ such that $\id'(u^*)=k$, and that 
there exists a node $v'\in V(G')$ such that $\certif_{id}(v)=\id'(v')$. Moreover, node $v$ verifies that the input $\inp'$ at $v'$ contains a collection of sets $\cS'(v')$ that contains $S'$, that is, $S'\in \cS'(v')$.
Finally, each node $v$ verifies that its immediate neighborhood $B_G(v,1)$ agrees with the corresponding neighborhood of $v'$ in $G'$, and that the candidate identities $\certif_{id}(w)$ of its neighbors  $w\in B_G(v,1)$ are compatible with the corresponding identities $\id'(w')$ in  $G'$. We term this verification the {\em neighborhood check} of $v$. 

It is easy to see  that when applying Algorithm  $A$ on a correct instance, together with the certificate described above, each node outputs ``yes''.
We now show the other direction. Assume  that Algorithm  $A$ applied on some triplet $(G,\inp,\certif)$ outputs ``yes'' at each node, our goal is to show that $(G,\inp)\in\cL$. Since all nodes say ``yes'' on $(G,\inp,\certif)$, it follows that the certificate $\certif(v)$ at every node $v\in V(G)$ contains the same candidate configuration field 
$(G',\inp',\id')$, the same candidate set $S'$ and the same pointer $\id'(v')$ to a vertex $v'\in G'$, such that 
$S'\in\cS(v')$. Since each node $v\in V(G)$ verifies  that $\cE(v)\in S'$, it follows that $S'\supseteq\{ \cE(v) \mid v\in V(G)\}$. It remains to show that there exists a node $v^*\in V(G)$ such that $S'\in\cS(v^*)$.  Indeed, this follows by the neighborhood checks of all nodes.
\end{proof}

\section{Future work}
This paper aims to make a first step in the direction of establishing a complexity theory for the locality discipline. 
Many interesting questions are left open. For example, it would be interesting to investigate the connections between 
$\BPLD(t,p,q)$ for different $p$ and $q$ such that $p^2+q\leq 1$. (A simple observation shows that $\BPLD(t,p,q)\subseteq \BPLD(t,p^k,1-(1-q)^k)$, for every integer $k$. Indeed, given an algorithm with a ``yes'' and ``no'' success probabilities $p$ and $q$, one can modify the success probabilities by performing 
$k$ runs and requiring each node to individually output ``no'' if it 
decided ``no'' on at least one of the runs. In this case, the ``no'' success 
probability increases from $q$ to at least $1-(1-q)^k$, and the ``yes'' 
success probability then decreases from $p$ to $p^k$.)   
Another interesting question is whether the phenomena we observed regarding randomization  occurs also in the 
non-deterministic setting, that is, 
whether  BPNLD$(t,p,q)$ collapses into  $\NLD(O(t))$, for $p^2+q>1$.

Our model of computation, namely, the $\cal{LOCAL}$ model, focuses on difficulties arising from purely locality issues, and abstracts away
other complexity measures. Naturally, it would be very interesting to come up with a rigorous complexity framework taking into account also other complexity measures.
For example, it would be interesting to investigate the connections between
classical computational complexity theory and the local complexity one. The bound on the (centralized) running time in each round (given by the function $f$, see
Section \ref{s:2}) may
serve a bridge for connecting the two theories, by putting constrains on this bound (i.e., $f$ must be polynomial, exponential, etc). 
Also, one could restrict the memory used by a node, in addition to, or instead of, bounding the sequential time.  Finally, it would be interesting  to come up with a complexity framework
taking also congestion into account.

\clearpage
\def\thepage{}

\commful
\newpage
\appendix
\centerline{\Large APPENDIX}

\parskip0.1in

\section{Proof of Claim \ref{claim:i}}\label{App:claim:closed}

\APPENDIXA
\qed


\section{Proof of Theorem \ref{thm:one-sided}}\label{App:one-sided}

\APPENDIXB
\qed

\section{Proof of Theorem \ref{theo:all}}\label{App:theo}

\APPENDIXC
\qed 

\section{Proof of Theorem \ref{separate-LD}}\label{App:separate-LD}

\APPENDIXD
\qed

\section{Proof of Theorem \ref{separate-NLD}}\label{App:separate-NLD}

\APPENDIXE
\qed 

\section{Proving $\mapcover\in \NLD$}\label{App:NLD}

\APPENDIXF
\qed

\section{Proof of Theorem \ref{theo:hardness2}}\label{App:theo:hardness2}

\APPENDIXG
\qed

\clearpage

\FIGA

\commfulend

\end{document}